\documentclass[11pt,a4paper]{article}
\usepackage{fullpage,graphicx}

\newcommand{\eBWT}{\ensuremath{\mathsf{eBWT}}}

\begin{document}

\title{MARIA: Multiple-alignment $r$-index with aggregation}
\author{Adri\'an Goga, Andrej Bal\'a\v{z}, Alessia Petescia and Travis Gagie}
\maketitle

\begin{abstract}
There now exist compact indexes that can efficiently list all the occurrences of a pattern in a dataset consisting of thousands of genomes, or even all the occurrences of all the pattern's maximal exact matches (MEMs) with respect to the dataset.  Unless we are lucky and the pattern is specific to only a few genomes, however, we could be swamped by hundreds of matches --- or even hundreds per MEM --- only to discover that most or all of the matches are to substrings that occupy the same few columns in a multiple alignment.  To address this issue, in this paper we present a simple and compact data index MARIA that stores a multiple alignment such that, given the position of one match of a pattern (or a MEM or other substring of a pattern) and its length, we can quickly list all the distinct columns of the multiple alignment where matches start.
\end{abstract}

\section{Introduction}

There now exist compact indexes that can efficiently list all the occurrences a pattern in a dataset consisting of thousands of genomes, such as the $r$-index~\cite{gagie2020fully,boucher2019prefix,kuhnle2020efficient}, or even all the occurrences of all the pattern's maximal exact matches (MEMs) with respect to the dataset~\cite{rossi2022moni}.  Unless we are lucky and the pattern is specific to only a few genomes, however, we could be swamped by hundreds of matches --- or even hundreds per MEM --- only to discover that most or all of the matches are to substrings that occupy the same few columns in a multiple alignment.

As a very small example of this, consider the toy multiple alignment shown in Figure~\ref{fig:alignment}.  Given the pattern {\sf ATT}, the indexes mentioned above will return all 3 occurrences, even though those all start in column 2 of the multiple alignment; given {\sf AT}, they will return all 8 occurrences even though those all start in columns 2 and 7.  It would be nice to be able to list all the distinct columns quickly, without enumerating the matches.  This idea is something like document listing --- which has already been applied in metagenomics and pangenomics~\cite{cobas2020tailoring} --- but that essentially summarizes the strings (the rows) containing matches, whereas here we are interested in the columns.  We refer to reporting only the distinct columns containing matches as {\em aggregation}.

We can augment the indexes cited above to support aggregation by adding grammar-compressed differential encodings of longest common prefix (LCP) arrays to find the suffix array (SA) intervals for patterns (see~\cite{gagie2020fully} for details) and then query something like augmented grammar-compressed document arrays containing column IDs --- but that approach has never been implemented (despite recent work suggesting it is not completely impractical).

\begin{figure}[t]
\begin{center}
\begin{tabular}{r|cccccccccc}
  &     0 &     1 &     2 &     3 &     4 &     5 &     6 &     7 &     8 &     9\\
\hline
0 & \tt - & \tt G & \tt A & \tt T & \tt T & \tt A & \tt C & \tt A & \tt T & \tt -\\
1 & \tt A & \tt G & \tt A & \tt T & \tt - & \tt A & \tt C & \tt A & \tt T & \tt -\\
2 & \tt - & \tt G & \tt A & \tt T & \tt - & \tt A & \tt C & \tt A & \tt T & \tt -\\
3 & \tt - & \tt G & \tt A & \tt T & \tt T & \tt A & \tt G & \tt A & \tt T & \tt -\\
4 & \tt - & \tt G & \tt A & \tt T & \tt T & \tt A & \tt G& \tt  A & \tt T & \tt A
\end{tabular}
\caption{A toy multiple alignment.}
\label{fig:alignment}
\end{center}
\end{figure}

Since computing the SA intervals is so daunting, in this paper we propose a workaround.  We consider run-length compressing a string {\sf col} containing the IDs of the columns where the row suffixes of the multiple alignment start, in their lexicographic order.  Although we cannot bound the number of runs $r'$ in {\sf col} in terms of, say, the number of runs $r$ in the extended Burrows-Wheeler Transform (eBWT) of the strings in the multiple alignment, we expect $r'$ will be small in practice: intuitively, the eBWT is usually run-length compressible because nearly all the columns in nearly all multiple alignments are nearly unary, so characters in the same columns are gathered together in the eBWT and form long runs --- and this should make {\tt col} compressible too.

We describe a simple index supporting aggregation based on run-length compressed {\sf col}.  Although our experiments are not yet complete, we are confident enough to call our index a multiple-alignment $r$-index with aggregation, or MARIA for short.  With MARIA, given the position of one match of a pattern (or a MEM or other substring of a pattern) and its length, we can quickly list all the distinct columns of the multiple alignment where matches start.  We are continuing our experiments and, when they are complete, we will consider how column IDs can be replaced by, for example, vertex IDs in a pangenome graph or marker IDs~\cite{DBLP:conf/wabi/MunKL22} (in which case {\sf col} can stand for ``colour'', which is often used to mean a generic classification).

\section{Data structure}

Given a multiple alignment, we first build the extended Burrows-Wheeler Transform (eBWT)~\cite{mantaci2007extension} of the aligned strings and then --- considering the rows to be cyclic but terminated by {\tt \$}s, and ignoring gaps --- we annotate each character $\eBWT [i]$ with the row and column of the multiple alignment where the suffix following $\eBWT [i]$ starts.  Figure~\ref{fig:eBWT} shows the annotated eBWT for our example.  Notice that, even though the first {\tt T} in row 1 is in column 3, in the eBWT it is annotated with a 5 for {\sf col} because there is a gap in column 4 and so the suffix {\tt ACAT\$AGA} that follows it starts in column 5.  The {\tt ?}s in {\sf col} indicate that there is no column of the multiple alignment containing {\tt \$}s.

\begin{figure}[t]
\begin{center}
\resizebox{\textwidth}{!}
{\begin{tabular}{c@{\hspace{3ex}}c@{\hspace{3ex}}c}
\begin{tabular}{cccl}
\sf row & \sf col & \sf eBWT & suffix\\
\hline
1 & ? & \tt T  & \tt \$AGATACA\\
2 & ? & \tt T  & \tt \$GATACA\\
0 & ? & \tt T  & \tt \$GATTACA\\
3 & ? & \tt T  & \tt \$GATTAGA\\
4 & ? & \tt A  & \tt \$GATTAGAT\\
4 & 9 & \tt T  & \tt A\$GATTAGA\\
1 & 5 & \tt T  & \tt ACAT\$AGA\\
2 & 5 & \tt T  & \tt ACAT\$GA\\
0 & 5 & \tt T  & \tt ACAT\$GAT\\
3 & 5 & \tt T  & \tt AGAT\$GAT\\
4 & 5 & \tt T  & \tt AGATA\$GAT\\
1 & 0 & \tt \$ & \tt AGATACAT\\
1 & 7 & \tt C  & \tt AT\$AGATA\\
2 & 7 & \tt C  & \tt AT\$GATA\\
0 & 7 & \tt C  & \tt AT\$GATTA
\end{tabular}
&
\begin{tabular}{cccl}
\sf row & \sf col & \sf eBWT & suffix\\
\hline
3 & 7 & \tt G  & \tt AT\$GATTA\\
4 & 7 & \tt G  & \tt ATA\$GATTA\\
1 & 2 & \tt G  & \tt ATACAT\$A\\
2 & 2 & \tt G  & \tt ATACAT\$\\
0 & 2 & \tt G  & \tt ATTACAT\$\\
3 & 2 & \tt G  & \tt ATTAGAT\$\\
4 & 2 & \tt G  & \tt ATTAGATA\$\\
1 & 6 & \tt A  & \tt CAT\$AGAT\\
2 & 6 & \tt A  & \tt CAT\$GAT\\
0 & 6 & \tt A  & \tt CAT\$GATT\\
3 & 6 & \tt A  & \tt GAT\$GATT\\
4 & 6 & \tt A  & \tt GATA\$GATT\\
1 & 1 & \tt A  & \tt GATACAT\$\\
2 & 1 & \tt \$ & \tt GATACAT\\
0 & 1 & \tt \$ & \tt GATTACAT
\end{tabular}
&
\begin{tabular}{cccl}
\sf row & \sf col & eBWT & suffix\\
\hline
3 & 1 & \tt \$ & \tt GATTAGAT\\
4 & 1 & \tt \$ & \tt GATTAGATA\\
1 & 8 & \tt A  & \tt T\$AGATAC\\
2 & 8 & \tt A  & \tt T\$GATAC\\
0 & 8 & \tt A  & \tt T\$GATTAC\\
3 & 8 & \tt A  & \tt T\$GATTAG\\
4 & 8 & \tt A  & \tt TA\$GATTAG\\
1 & 3 & \tt A  & \tt TACAT\$AG\\
2 & 3 & \tt A  & \tt TACAT\$G\\
0 & 4 & \tt T  & \tt TACAT\$GA\\
3 & 4 & \tt T  & \tt TAGAT\$GA\\
4 & 4 & \tt T  & \tt TAGATA\$GA\\
0 & 3 & \tt A  & \tt TTACAT\$G\\
3 & 3 & \tt A  & \tt TTAGAT\$G\\
4 & 3 & \tt A  & \tt TTAGATA\$G
\end{tabular}
\end{tabular}}
\caption{The eBWT for the strings in Figure~\ref{fig:alignment} together with the suffix following each character in its string and the row and column in the multiple alignment where that suffix starts.}
\label{fig:eBWT}
\end{center}
\end{figure}

For each run in the string {\sf col} storing the column annotations, we store the run-head (that is, the distinct entry in that run) and the top and bottom row annotations for that run; we then discard the eBWT (and will not use any further BWT-based techniques).  Figure~\ref{fig:MARIA} shows what we store for our example.

We also store a data structure supporting fast longest common extension (LCE) queries: given two triples $(i, j)$ and $(i', j')$, it returns the length of the longest common prefix of the suffixes starting in row $i$ and column $j$ and in row $i'$ and column $j'$ of the multiple alignment, and indicates which of those suffixes is lexicographically smaller (again considering rows to be cyclic when necessary).  For our example, if $(i, j) = (0, 7)$ and $(i', j') = (3, 2)$, then the LCE data structure returns $(2, \prec)$, because the suffix {\tt AT\$GATTAC} starting in row 0 and column 7 and the suffix {\tt ATTAGAT\$G} starting in row 3 and column 2 have a longest common prefix of length 2 and the former suffix is lexicograpically smaller; if $(i, j) = (3, 4)$ and $(i', j') = (2, 8)$ then the LCE data structure returns $(1, \succ)$.  Compact LCE data structures are common in the literature (see, e.g.,~\cite{boucher2021phoni}) and, although most do not compare the strings by default, it is easy to modify them to do so.  We leave the details for the full version of this paper.

Finally, we store an $O (r')$-space data structure --- where $r'$ is again the number of runs in {\sf col} --- that, given an interval in our string {\sf heads} of {\sf col} run-heads, lists the distinct entries in that interval in constant time per entry listed.  Muthukrishnan~\cite{muthukrishnan2002efficient} described the first data structure for this problem and there are now several other options (see, e.g.,~\cite{gagie2013colored}).  Again, we leave the details for the full version of this paper.

In total we use $O (r')$ space plus that for the LCE data structure, which is small in practice.

\begin{figure}[t]
\begin{center}
\begin{tabular}{ccc}
\sf heads & \sf t-row & \sf b-row\\
\hline
? & 1 & 4\\
9 & 4 & 4\\
5 & 1 & 4\\
0 & 1 & 1\\
7 & 1 & 4\\
2 & 1 & 4\\
6 & 1 & 4\\
1 & 1 & 4\\
8 & 1 & 4\\
3 & 1 & 2\\
4 & 0 & 4\\
3 & 0 & 4
\end{tabular}
\caption{The basic MARIA index for our example.  The data structures for LCE queries and listing the distinct elements in intervals of {\sf col} are fairly standard and thus not shown.}
\label{fig:MARIA}
\end{center}
\end{figure}

\section{Queries}

Given the position of one match of a pattern (or a MEM or other substring of a pattern) as a pair $(i, j)$ indicating the row and column of the multiple alignment where the match starts, and the length $\ell$ of the match, we can first use the LCE data structure and binary search
\begin{enumerate}
\item to find the entry in {\sf heads} for the run in {\sf col} that would contain the column annotation $j$ for the character annotated $(i, j)$ in the eBWT,
\item to find the interval in {\sf heads} containing entries from the runs in {\sf col} that would column annotations $j'$ for all the characters annotated $(i', j')$ such that the LCE of $(i, j)$ and $(i', j')$ is at least $\ell$.
\end{enumerate}
We emphasize that MARIA will work whether the values $i$, $j$ and $\ell$ come from a BWT-based data structure or from some completely different source.

Suppose that, for our example, we are given the starting position $(3, 4)$ of an occurrence of the match {\tt TA} and its length 2.  In the first step of the first binary search, we use the LCE data structure to compare $(3, 4)$ to $(4, 2)$ --- the $(i', j')$ pair for the bottom row of the {\sf col} run that starts with annotation $(1, 2)$ and ends with $(4, 2)$, stored in MARIA as the row $2, 1, 4$ --- and it returns $(0, \succ)$, meaning the suffix starting in row 3 and column 4 ({\tt TAGAT\$GAT}) is larger than the one ({\tt ATTAGATA\$G}) starting in row 4 and column 2.  In the second step, we use the LCE data structure to compare $(3, 4)$ to $(4, 8)$ --- the $(i', j')$ pair for the bottom row of the {\sf col} run that starts with the annotation $(1, 8)$ and ends with $(4, 8)$, stored in MARIA as the row $8, 1, 4$ --- and it returns $(2, \succ)$.  We keep going like this, comparing $(3, 4)$ against $(0, 4)$ and receiving $(2, \succ)$ and then against $(0, 3)$ and receiving $(1, \prec)$ and finally against $(4, 4)$ and receiving $(5, \prec)$, and concluding that the annotation $(3, 4)$ is in the {\sf col} run that starts with annotation $(0, 4)$ and ends with annotation $(4, 4)$.  We admit this is not the most exciting example imaginable, but it is at least manageable.

Once we know the 10th {\sf col} run (counting from 0) contains the annotation $(3, 4)$, we can use binary search to determine that the range of annotations $(i', j')$ such that the LCE for $(3, 4)$ and $(i', j')$ is at least 2, starts after the first annotation $(1, 8)$ in the 8th {\sf col} run but at or before the last annotation $(4, 8)$ in that run, and ends before the first $(0, 3)$ in the 11th {\sf col} run.  Therefore, we report the distinct entries between the 8th and 10th runs, which are columns 8, 3 and 4 --- which are indeed where occurrences of {\tt TA} start in the multiple alignment.  In this case all the entries in $\mathsf{heads} [8..10]$ are unique, but if we repeat the search for $(3, 4)$ with length 1 instead of 2, then we end up reporting only the 3 distinct entries in the range ${\sf heads} [8..11]$ of length 4.

Assuming the LCE data structure works in $O (\log n)$ time, where $n \geq r'$ is total number of cells in the multiple alignment, the binary searches take $O (\log^2 n)$ time, and then reporting the distinct columns takes $O (1)$ time per column.  We can combine the searches into one and speed that up by pre-computing and storing certain LCE values and keeping data structures supporting range-minimum queries (RMQs), still using $O (r')$ space plus that for the LCE data structure, but we again leave the details for the full version of this paper.  If we find the starting location $(i, j)$ of the match with a BWT-based data structure and it also gives us the lexicographic rank of the suffix starting at row $i$ and column $j$, then we can skip the first binary search and replace the second with 2 LCE queries and RMQs, so we use $O (\log n)$ total time.

\subsection*{Acknowledgements}

Many thanks to M\'aria Gogov\'a for moral support; to Uwe Baier, Christina Boucher, Bro\v{n}a Brejov\'a, Ben Langmead, Taher Mun, Gonzalo Navarro, Max Rossi, Jouni Sir\'en and Tom\'a\v{s} Vina\v{r} for helpful discussions; to the organizers and other participants of the Lipari workshop ``The future of compressed data structures'' and the ALPACA / PANGAIA pangenomics workshop after WABI 2022; and to Pinar Bistro in Nuthetal for kebabs and a table where we worked out the ideas in this paper.  TG is funded by NSERC grant RGPIN-07185-2020 and subawards from NIH grant HG011392 and NSF grant 2029552.

\bibliographystyle{plain}
\bibliography{MARIA}

\end{document}